\documentclass[conference]{IEEEtran}
\IEEEoverridecommandlockouts
\usepackage{cite}
\usepackage{amsmath,amssymb,amsfonts}
\usepackage{algorithmic}
\usepackage{graphicx}
\usepackage{textcomp}
\usepackage{xcolor}
\usepackage{amsmath} 
\def\BibTeX{{\rm B\kern-.05em{\sc i\kern-.025em b}\kern-.08em
    T\kern-.1667em\lower.7ex\hbox{E}\kern-.125emX}}
\begin{document}

\title{Securing Majority-Attack In Blockchain Using Machine Learning And Algorithmic Game Theory: A Proof of Work\\
\thanks{* This work was supported in part by a grant from Codeepy Pvt. Ltd. with reference CDPY/2018/1.}
}

\author{\IEEEauthorblockN{Somdip Dey}
\IEEEauthorblockA{\textit{School of Computer Science and Electronic Engineering} \\
\textit{University of Essex}\\
Colchester, UK \\
somdip.dey@essex.ac.uk}
}

\maketitle

\begin{abstract}
Recently we could see several institutions coming together to create consortium based blockchain networks such as Hyperledger. Although for applications of blockchain such as Bitcoin, Litcoin, etc. the majority-attack might not be a great threat but for consortium based blockchain networks where we could see several institutions such as public, private, government, etc. are collaborating, the majority-attack might just prove to be a prevalent threat if collusion among these institutions takes place. This paper proposes a methodology where we can use intelligent software agents to monitor the activity of stakeholders in the blockchain networks to detect anomaly such as collusion, using supervised machine learning algorithm and algorithmic game theory and stop the majority-attack from taking place.\\
\end{abstract}


\section{Introduction}
When Satoshi Nakamoto [2] released the technology named Bitcoin, he revolutionised the industry not because he has invented a new currency system, which do not require intervention of institutional mediator while transferring money from one entity to another, but because he has gifted one of the most disruptive technology, which has come to life in decades. With the introduction of Bitcoin, Blockchain got introduced to the world, which is a digital ledger in which all transactions are recorded chronologically and publicly. But the application of blockchain is not just limited to crypto-currencies [3, 4] such as Bitcoin and have proved to be useful in tracking ownership, provenance of documents, digital assets, physical assets, voting rights, etc. As we can see the rise in use of blockchain technologies, we can also see rise of security issues such as 'Double-Spending', especially in the Majority Attack [2, 5, 6, 7, 8]. The majority attack is carried out by a group of individuals/entities in the decentralised environment, who colludes to take control over the ledger to gain profit from it. In this proposed methodology, a novel approach of using Algorithmic Game Theory concepts and Machine Learning techniques is used to reduce the chances of collusion in the decentralized system to gain advantage over other miners so that the system can be as fair as possible.
In section 2, some background theory regarding blockchain and economy of double-spending is discussed. In section 3, the proposed methodology is discussed. Finally the paper ends with some discussion in section 4 and conclusion.

\section{Background Theory And Related Work}
In this section we will visit the concepts of double-spending and the majority attack in Blockchain. Later in this section we will also discuss the economy of the attack being performed and how Game Theory is applicable to security implications in blockchain as well. 

\subsection{Blockchain, Double-Spending \& Majority Attack}
A double-spending attack [2, 5, 6, 7] in blockchain means the attacker has to convince the merchant that a transaction has been confirmed and then convince the entire network to approve some other transaction, which will lead to the attacker keeping both the money and the service (goods) from the merchant whereas the merchant would be left with neither the money or the service. This problem in synchronization is solved by proof-of-work, which is a computational effort consisting of hashes to acknowledge the groups of transactions, also known as blocks. For a transaction to be valid, sufficient work has been done to acknowledge that the block contains it. Since, validation of blocks require computational effort to do so, this also gives rise to another issue, what if the attacker has substantial computational power at its disposal? All the attacker has to do is mine a blockchain privately till the length of the chain becomes longer than the chain mined by the honest network, and release this private blockchain for confirmation when it is appropriate. 

In Rosenfeld's paper [6] the probability of the attacker succeeding in his attack is discussed. If we consider \textit{z} as the number of blocks by which the honest network has advantage over the attacker then \textit{z = n - m}, where \textit{n} is the number of blocks in the chain on top of the one where fork started for the honest network, whereas m is the number of blocks in the chain on top of the fork which the attacker has built. Before we discuss the probability of having advantage over the attacker, let us consider the following assumptions:
\begin{itemize}
\item The total hashrate of the attacker and honest network is constant. They have a hashrate of \textit{H} combined, of which \textit{pH} belongs to honest network and \textit{qH} belongs to the attacker, where \textit{p + q = 1}.
\item The mining difficulty is constant, such that the time taken to find a block with \textit{H} hashrate is \textit{T\textsubscript{o}}.
\end{itemize}

There are two possibilities of double-spending attack, which is either the attack succeeds or it fails, as follows:

\begin{equation*}
z_{i+1}=\begin{cases}
z_i+1, & \text{\textit{with probability p}}.\\
z_i-1, & \text{\textit{with probability q}}.
\end{cases}
\end{equation*}

If we consider \textit{a\textsubscript{z}} to be the probability of the attacker succeeding in the attack then we can arrive at the following equation:

\begin{equation}
a_z=pa_{z+1}+qa_{z-1}
\end{equation}

And if we solve this using the boundary condition and the notion \textit{p + q = 1} then we can conclude:

\begin{equation}
\begin{split}
a_z = \min({\frac{q}{p},1})^{\max({z+1,0})}\\
 =\begin{cases}
1, & if z<0 || q>p\\
〖(\frac{q}{p})^{z+1}, & if z▁(>) 0 || q▁(<) p
\end{cases}
\end{split}
\end{equation}

If we assume \textit{n} number of blocks are found by the honest network and \textit{m + 1} number of blocks are found by the attacker during this time period then the probability (\textit{r}) of double-spending to succeed when the merchant waits for \textit{n} confirmations using the equation (2) is:

\begin{equation}
\begin{split}
r=\sum_{m=0}^{\infty}P〖(m)a〗_{(n-m-1)}\\
 =\begin{cases}
 1 - \sum_{m=0}^{n}(\begin{array}{c}
 m+n-1\\m
 \end{array}) (p^n q^m- p^m q^n ) & if q<p\\
 1 & if q>p
 \end{cases}
\end{split}
\end{equation}

In the study [6], it is proved that as the number of confirmations by the honest network increased, the success rate of the attack decreased but no matter how many confirmations by the honest network has succeeded, the attack will always succeed if the hashrate of the attacker approached 50\% of the total network hashrate, which means \textit{q} $\ge  0.5 $. 

This proves that an attacker with more computing power at its disposal might prove to be a key factor in succeeding in the attack. This particularly raises security concerns in Consortium Blockchain [5, 7, 8] such as Hyperledger, where we can see involvement of several companies or business entities. Whoever in the Hyperledger network holds the maximum computing power, can always get a competitive advantage over its competitors while performing business transaction over the network. 

With Proof of Work, more CPU/GPU power is required in checking hashes of each block in the blockchain. Because of this mechanism, more and more business entities would like to join in this mining process, which would create ?mining pools?, and once the mining pool holds 51\% computing power, then it would take control of the blockchain.  Therefore, by taking control what it can do is [5, 8]:

\begin{itemize}
\item Modify the transaction data, which can lead to double spending attack
\item To stop the block verification transaction
\item To stop miners mining any available block
\end{itemize}

\subsection{Economy of Double-Spending}
In the study by Rosenfeld [6], it was found that the number of confirmations required to keep the success rate of the attacker (double-spending) below 10\%, 1\% and 0.1\%, are 2, 4 and 6 respectively. In addition, we have already seen that once the attacker's hashrate reaches 50\% of the total network hashrate then the number of confirmations required reaches infinity, which means no amount of confirmation can defeat the attack. Taking this into account, we also have to consider the likelihood of the attack being performed in reality. If value of the commodity being exchanged is assumed to have a value of \textit{v} and the attacker has mined \textit{o} number of blocks where each block has a value of \textit{B}, then if the attack succeeds the attacker will gain \textit{v}, where if the attack fails then the attacker will loose \textit{v + oB}. Therefore, if we consider the two possibilities, the payoff (\textit{s}) for the attacker is as follows:

\begin{equation}
\begin{split}
s =\begin{cases}
v, & if q => 0.5\\
-(v+oB), &   if q<0.5
\end{cases}
\end{split}
\end{equation}

\begin{equation*}
\text{\textit{where q is the hashrate of the attacker}}
\end{equation*}

And in order to carry on with the attack the value of \textit{v} has to be significant. This payoff (\textit{s}) will prove to be useful in portraying the security implication in the light of Game Theory, and how decisions can be made to classify whether an attack is taking place or not.

\section{Proposed Methodology}
In section 2.B we have already seen that payoff (\textit{s}) for the attacker can only have two possibilities: succeed or fail. This is where Game Theory [1] comes into account. But before we get into the concept let us define few terminologies of Game Theory in this context as follows:

\begin{itemize}
\item \textit{Self-Interested Agents}: This can be any entity such as a person, business or any other institution in the blockchain network with their own preferences and utility. This also includes honest entities and attacker(s).
\item \textit{Player}: Each Self-Interested Agent who are participating in the blockchain network. Let us assume that there are N players where N = (1,?.., n) is a finite set of n, indexed by i.
\item \textit{Action}: Action taken by each player based on their preferences and utility. And let us assume that set of actions taken by the player i is Ai where Ai = (a1,?., an)
\item \textit{Payoff}: The reward, which each player receives
\end{itemize}

Now, if we consider the equation (4) then we can see the attacker would want to maximize the probability of getting a payoff of \textit{v} instead of loosing \textit{v + oB}. Therefore, we can extend the same equation (4) to derive the utility function/ payoff function as follows:

\begin{equation}
\begin{split}
u(a) =\begin{cases}
v, & if q => 0.5\\
-(v+oB), &   if q<0.5
\end{cases}
\end{split}
\end{equation}

where \textit{u} is utility, \textit{a} is the action taken by the attacker, \textit{q} is the hashrate of the attacker, \textit{v} is value of commodity/service by the merchant, \textit{o} is number of blocks mined, \textit{B} is value of each block.

This utility function (\textit{v}) will govern the decision on whether an attack is bound to happen or not by the attacker based on the value of the commodity/service. And in order to keep the blockchain network safe from the Majority Attack we should focus on this function. 

We can feed this utility function to Supervised Machine Learning algorithms to classify whether an attack is likely to take place or not. If the attack is likely to take place then set of rules should be implemented by the system to either prevent the blockchain confirmation from the attacker(s) or to prevent confirmation of the whole transaction till a new fair transaction is performed again i.e. no payoffs for anyone, in order to ensure fairness and legitimate transactions being confirmed in the network. 

In order to achieve this, an intelligent agent is implemented in the application layer of the blockchain network system, which would have two distinct parts:
\begin{enumerate}
\item Based on the past transactions of the stakeholders the probability of each stakeholder to defect
\item Based on the current value of the commodity/service being sold in the current transaction the probability of the stakeholder(s) to attack through majority attack
\end{enumerate}

\begin{figure}[!h]
\centering
\includegraphics[scale=0.52]{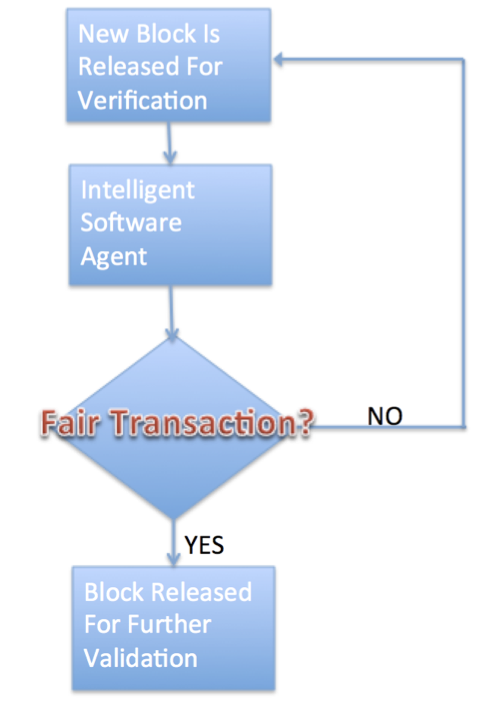}
\caption{Proof-of-Work (The Proposed Methodology)}
\label{fig1}
\vspace{-2mm}
\end{figure}

In Fig.\ref{fig1}, we can see the workflow of the proposed methodology where after the new block is released in the network for the consensus purpose by the stakeholders (including attacker and other players), the intelligent software agent in the application layer of the network uses the utility function (\textit{v}) to classify the motive of the stakeholders and the value of the current service being sold in the transaction. If the motive of the stakeholders is deemed to be malicious in nature with the intent of collusion to perform a majority attack then the transaction is cancelled and all the stakeholders are requested for a new transaction instead.

\section{Discussion}
In the proposed methodology we have discussed about the utility function (\textit{v}) based on the value of the service or commodity being sold in the current transaction. Here, the commodity does not have to be something that has a tangible value in the network, rather it can have some personal attachment or importance to the stakeholder(s). In that case the intelligent agent needs to deduce the level of attachment or importance of the commodity or service being handled in the transaction in order to calculate the utility function and then the probability of the majority attack from taking place.

\section{Conclusion}
As blockchain technology becomes more and more popular, we can see emergence of several variations of such consensus based distributed ledger systems where majority-attack can become more proficient. In order to prevent such malicious activity in the consensus based distributed ledger systems we can utilise some variations of the Proof-of-Work proposed in this paper. Although this is a work in progress and in it?s preliminary stage, the proposed Proof-of-Work will be extended to provide more holistic approach to such issues faced in the system.

\section*{Acknowledgment}

The author wishes to thank his colleagues at ReMe Basket Ltd. and Codeepy UK Pvt. Ltd. for their support. He would also like to thank his parents, Soma Dey and Sudip Dey, for their continued support and faith on the author?s capabilities. This work was supported in part by a grant from Codeepy Pvt. Ltd. with reference CDPY/2018/1.

\vspace{12pt}

\end{document}